\documentstyle[twoside,psfig]{article}

\catcode`\@=11
\long\def\@makefntext#1{
\protect\noindent \hbox to 3.2pt {\hskip-.9pt  
$^{{\eightrm\@thefnmark}}$\hfil}#1\hfill}		

\def\thefootnote{\fnsymbol{footnote}}
\def\@makefnmark{\hbox to 0pt{$^{\@thefnmark}$\hss}}	
	
\def\ps@myheadings{\let\@mkboth\@gobbletwo
\def\@oddhead{\hbox{}
\rightmark\hfil\eightrm\thepage}   
\def\@oddfoot{}\def\@evenhead{\eightrm\thepage\hfil
\leftmark\hbox{}}\def\@evenfoot{}
\def\sectionmark##1{}\def\subsectionmark##1{}}

\oddsidemargin=\evensidemargin
\renewcommand{\thefootnote}{\fnsymbol{footnote}}

\newcounter{sectionc}\newcounter{subsectionc}\newcounter{subsubsectionc}
\renewcommand{\section}[1] {\vspace{12pt}\addtocounter{sectionc}{1} 
\setcounter{subsectionc}{0}\setcounter{subsubsectionc}{0}\noindent 
	{\tenbf\thesectionc. #1}\par\vspace{5pt}}
\renewcommand{\subsection}[1] {\vspace{12pt}\addtocounter{subsectionc}{1} 
	\setcounter{subsubsectionc}{0}\noindent 
	{\bf\thesectionc.\thesubsectionc. {\kern1pt \bfit #1}}\par\vspace{5pt}}
\renewcommand{\subsubsection}[1] {\vspace{12pt}\addtocounter{subsubsectionc}{1}
	\noindent{\tenrm\thesectionc.\thesubsectionc.\thesubsubsectionc.
	{\kern1pt \tenit #1}}\par\vspace{5pt}}
\newcommand{\nonumsection}[1] {\vspace{12pt}\noindent{\tenbf #1}
	\par\vspace{5pt}}

\newcounter{appendixc}
\newcounter{subappendixc}[appendixc]
\newcounter{subsubappendixc}[subappendixc]
\renewcommand{\thesubappendixc}{\Alph{appendixc}.\arabic{subappendixc}}
\renewcommand{\thesubsubappendixc}
	{\Alph{appendixc}.\arabic{subappendixc}.\arabic{subsubappendixc}}

\renewcommand{\appendix}[1] {\vspace{12pt}
        \refstepcounter{appendixc}
        \setcounter{figure}{0}
        \setcounter{table}{0}
        \setcounter{lemma}{0}
        \setcounter{theorem}{0}
        \setcounter{corollary}{0}
        \setcounter{definition}{0}
        \setcounter{equation}{0}
        \renewcommand{\thefigure}{\Alph{appendixc}.\arabic{figure}}
        \renewcommand{\thetable}{\Alph{appendixc}.\arabic{table}}
        \renewcommand{\theappendixc}{\Alph{appendixc}}
        \renewcommand{\thelemma}{\Alph{appendixc}.\arabic{lemma}}
        \renewcommand{\thetheorem}{\Alph{appendixc}.\arabic{theorem}}
        \renewcommand{\thedefinition}{\Alph{appendixc}.\arabic{definition}}
        \renewcommand{\thecorollary}{\Alph{appendixc}.\arabic{corollary}}
        \renewcommand{\theequation}{\Alph{appendixc}.\arabic{equation}}
        \noindent{\tenbf Appendix \theappendixc #1}\par\vspace{5pt}}
\newcommand{\subappendix}[1] {\vspace{12pt}
        \refstepcounter{subappendixc}
        \noindent{\bf Appendix \thesubappendixc. {\kern1pt \bfit #1}}
	\par\vspace{5pt}}
\newcommand{\subsubappendix}[1] {\vspace{12pt}
        \refstepcounter{subsubappendixc}
        \noindent{\rm Appendix \thesubsubappendixc. {\kern1pt \tenit #1}}
	\par\vspace{5pt}}

\newcommand{\textlineskip}{\baselineskip=13pt}
\newcommand{\smalllineskip}{\baselineskip=10pt}

\def\eightcirc{
\begin{picture}(0,0)
\put(4.4,1.8){\circle{6.5}}
\end{picture}}
\def\eightcopyright{\eightcirc\kern2.7pt\hbox{\eightrm c}} 

\newcommand{\copyrightheading}[1]
	{\vspace*{-2.5cm}\smalllineskip{\flushleft
	{\footnotesize International Journal of Modern Physics A #1}\\
	{\footnotesize $\eightcopyright$\, World Scientific Publishing
	 Company}\\
	 }}

\newcommand{\publisher}[2]{{\begin{center}\footnotesize\smalllineskip 
	Received #1\\
	Revised #2
	\end{center}
	}}

\def\abstracts#1#2#3{{
	\centering{\begin{minipage}{4.5in}\footnotesize\baselineskip=10pt
	\parindent=0pt #1\par 
	\parindent=15pt #2\par
	\parindent=15pt #3
	\end{minipage}}\par}} 

\newcommand{\bibit}{\nineit}

\renewenvironment{thebibliography}[1]
	{\frenchspacing
	 \ninerm\baselineskip=11pt
	 \begin{list}{\arabic{enumi}.}
	{\usecounter{enumi}\setlength{\parsep}{0pt}
	 \setlength{\leftmargin 12.7pt}{\rightmargin 0pt} 
	 \setlength{\itemsep}{0pt} \settowidth
	{\labelwidth}{#1.}\sloppy}}{\end{list}}

\newcounter{itemlistc}
\newcounter{romanlistc}
\newcounter{alphlistc}
\newcounter{arabiclistc}

\newcommand{\fcaption}[1]{
        \refstepcounter{figure}
        \setbox\@tempboxa = \hbox{\footnotesize Fig.~\thefigure. #1}
        \ifdim \wd\@tempboxa > 5in
           {\begin{center}
        \parbox{5in}{\footnotesize\smalllineskip Fig.~\thefigure. #1}
            \end{center}}
        \else
             {\begin{center}
             {\footnotesize Fig.~\thefigure. #1}
              \end{center}}
        \fi}

\newcommand{\tcaption}[1]{
        \refstepcounter{table}
        \setbox\@tempboxa = \hbox{\footnotesize Table~\thetable. #1}
        \ifdim \wd\@tempboxa > 5in
           {\begin{center}
        \parbox{5in}{\footnotesize\smalllineskip Table~\thetable. #1}
            \end{center}}
        \else
             {\begin{center}
             {\footnotesize Table~\thetable. #1}
              \end{center}}
        \fi}

\def\@citex[#1]#2{\if@filesw\immediate\write\@auxout
	{\string\citation{#2}}\fi
\def\@citea{}\@cite{\@for\@citeb:=#2\do
	{\@citea\def\@citea{,}\@ifundefined
	{b@\@citeb}{{\bf ?}\@warning
	{Citation `\@citeb' on page \thepage \space undefined}}
	{\csname b@\@citeb\endcsname}}}{#1}}

\newif\if@cghi
\def\cite{\@cghitrue\@ifnextchar [{\@tempswatrue
	\@citex}{\@tempswafalse\@citex[]}}
\def\citelow{\@cghifalse\@ifnextchar [{\@tempswatrue
	\@citex}{\@tempswafalse\@citex[]}}
\def\@cite#1#2{{$\null^{#1}$\if@tempswa\typeout
	{IJCGA warning: optional citation argument 
	ignored: `#2'} \fi}}

\def\pmb#1{\setbox0=\hbox{#1}
	\kern-.025em\copy0\kern-\wd0
	\kern.05em\copy0\kern-\wd0
	\kern-.025em\raise.0433em\box0}

\def\fnt#1#2{\footnotetext{\kern-.3em
	{$^{\mbox{\scriptsize #1}}$}{#2}}}

\def\thefootnote{\fnsymbol{footnote}}
\def\@makefnmark{\hbox to 0pt{$^{\@thefnmark}$\hss}}	
	
\def\ps@myheadings{%
    \let\@oddfoot\@empty\let\@evenfoot\@empty
    \def\@evenhead{\slshape\leftmark\hfil}
    \def\@oddhead{\hfil{\slshape\rightmark}}
    \let\@mkboth\@gobbletwo
    \let\sectionmark\@gobble
    \let\subsectionmark\@gobble
    }
\font\tenrm=cmr10
\font\tenit=cmti10 
\font\tenbf=cmbx10
\font\bfit=cmbxti10 at 10pt
\font\ninerm=cmr9
\font\nineit=cmti9

\font\eightrm=cmr8

\def\beqra{\begin{eqnarray}} \def\eeqra{\end{eqnarray}}
\def\beqast{\begin{eqnarray*}} \def\eeqast{\end{eqnarray*}}
\def\beq{\begin{equation}}      \def\eeq{\end{equation}}
\def\be{\begin{enumerate}}   \def\ee{\end{enumerate}}

\def\gam{\gamma}

\def\si{\sigma}

\def\del{\delta}

\def\om{\omega}

\def\cd{{\cal D}}

\def\cz{{\cal{Z}}}

\def\raisenot{\raise .5mm\hbox{/}}
\def\nota{\ \hbox{{$a$}\kern-.49em\hbox{/}}}
\def\notA{\hbox{{$A$}\kern-.54em\hbox{\raisenot}}}
\def\notb{\ \hbox{{$b$}\kern-.47em\hbox{/}}}
\def\notB{\ \hbox{{$B$}\kern-.60em\hbox{\raisenot}}}
\def\notc{\ \hbox{{$c$}\kern-.45em\hbox{/}}}
\def\notd{\ \hbox{{$d$}\kern-.53em\hbox{/}}}
\def\notbd{\ \hbox{{$D$}\kern-.61em\hbox{\raisenot}}} 
\def\note{\ \hbox{{$e$}\kern-.47em\hbox{/}}}
\def\notk{\ \hbox{{$k$}\kern-.51em\hbox{/}}}
\def\notp{\ \hbox{{$p$}\kern-.43em\hbox{/}}}
\def\notq{\ \hbox{{$q$}\kern-.47em\hbox{/}}}
\def\notW{\ \hbox{{$W$}\kern-.75em\hbox{\raisenot}}}
\def\notz{\ \hbox{{$Z$}\kern-.61em\hbox{\raisenot}}}
\def\notpa{\hbox{{$\partial$}\kern-.54em\hbox{\raisenot}}}
\def\fo{\hbox{{1}\kern-.25em\hbox{l}}}  

\def\tr{{\rm Tr}}

\def\sigx{\sigma(x)}

\def\qed{\hbox{${\vcenter{\vbox{			
   \hrule height 0.4pt\hbox{\vrule width 0.4pt height 6pt
   \kern5pt\vrule width 0.4pt}\hrule height 0.4pt}}}$}}

\renewcommand{\thefootnote}{\fnsymbol{footnote}}  

\begin{document}
\setlength{\textheight}{7.7truein}  

\thispagestyle{empty}

\markboth{\protect{\footnotesize\it Periodic Table of Fermion 
Bags}}{\protect{\footnotesize\it Periodic Table of Fermion Bags}}

\normalsize\textlineskip

\setcounter{page}{1}

\copyrightheading{}		

\vspace*{0.88truein}

\centerline{\bf THE PERIODIC TABLE OF STATIC FERMION BAGS}
\vspace*{0.035truein}
\centerline{\bf IN THE GROSS-NEVEU MODEL}
\vspace*{0.27truein}
\centerline{\footnotesize JOSHUA FEINBERG}
\baselineskip=12pt
\centerline{\footnotesize\it Physics Department, University
of Haifa at Oranim, Tivon 36006, Israel\footnote{Permanent address}}
\baselineskip=10pt
\centerline{\footnotesize\it and}
\baselineskip=10pt
\centerline{\footnotesize\it Physics Department, Technion, Haifa 32000,
Israel}
\vspace*{0.225truein}
\publisher{(received date)}{(revised date)}

\vspace*{0.21truein}
\abstracts{We study the spectrum of stable static fermion bags in the 
$1+1$ dimensional Gross-Neveu model with $N$ flavors of Dirac fermions, in 
the large $N$ limit. In the process, we discover a new kink, heavier than the 
Callan-Coleman-Gross-Zee (CCGZ) kink, which is marginally stable (at least 
in the large $N$ limit). The connection of this new kink and the conjectured 
$S$ matrix of the Gross-Neveu model is obscured at this point. 
After identifying all stable static fermion bags, we arrange them into a 
periodic table, according to their $O (2N)$ and topological quantum 
numbers.}{}{}


\vspace*{1pt}\textlineskip	
\section{Introduction}	        
\vspace*{-0.5pt}
\noindent
A central concept in particle physics states that
fundamental particles acquire their masses through interactions with vacuum
condensates. Thus, a massive particle may carve out around itself a spherical
region \cite{sphericalbag} or a shell \cite{shellbag} in which the
condensate is suppressed, thus reducing the effective mass of the particle
at the expense of volume and gradient energy associated with the
condensate. This picture has interesting phenomenological consequences
\cite{sphericalbag,mackenzie}.

This phenomenon may be studied non-perturbatively in model field theories 
in $1+1$ space-time dimensions such as the Gross-Neveu (GN) model \cite{gn}
and the multi-flavor Nambu-Jona-Lasinio (NJL) \cite{njl} model,
in the large $N$ limit. These models are particularly appealing, since they
exhibit, among other things, asymptotic freedom and dynamical mass 
generation, like more realistic four dimensional models.

Many years ago, Dashen, Hasslacher and Neveu (DHN) \cite{dhn}, and following 
them Shei \cite{shei}, used inverse scattering analysis \cite{faddeev} to find 
static fermion-bag \cite{sphericalbag,shellbag} soliton solutions to the 
large-$N$ saddle point equations of the GN \cite{gn} and of the 
$1+1$ dimensional, multi-flavor NJL \cite{njl} models.
In the GN model, with its discrete chiral symmetry, a topological soliton,
the so called Callan-Coleman-Gross-Zee (CCGZ) kink \cite{ccgz}, was 
discovered prior to the work of DHN. In this report we will concentrate 
exclusively on the GN model.

One version of writing the action of the $1+1$ dimensional GN model is 
$S=\int d^2x\,\left\{\sum_{a=1}^N\, \bar\psi_a\,\left(i\notpa-\si
\right)\,\psi_a 
-{1\over 2g^2}\,\si^2\right\}\,,$
where the $\psi_a\,(a=1,\ldots,N)$ are $N$ flavors of massless Dirac 
fermions, with Yukawa coupling to the scalar auxiliary field 
$\si(x)$. This action is evidently symmetric under 
the simultaneous transformations $\si\rightarrow -\si$ and 
$\psi\rightarrow\gam_5\psi$, which generate the so-called 
discrete (or ${\bf Z}\!\!\!{\bf Z}_2$) chiral symmetry of the GN model. 
The GN action has also flavor symmetry $O(2N)$, which can be seen by breaking
the $N$ Dirac spinors into $2N$ Majorana spinors. Related to this is the 
fact that the model is also invariant under charge-conjugation.

Performing functional integration over the grassmannian variables in 
the GN action leads to the partition function $
\cz=\int\,\cd\si\,\exp \{iS_{eff}[\si]\}$
where the bare effective action is
\beq
S_{eff}[\si] =-{1\over 2g^2}\int\, d^2x 
\,\si^2-iN\, 
\tr\log\left(i\notpa-\si\right)
\label{effective}
\eeq
and the trace is taken over both functional and Dirac indices.

The theory (\ref{effective}) has been studied in the limit 
$N\rightarrow\infty$ with $Ng^2$ held fixed\cite{gn}. In this limit 
the partition function $\cz$ is governed by saddle points of (\ref{effective}) 
and the small fluctuations around them. The most general saddle point
condition reads
\beqra
{\del S_{\em eff}\over \del \si\left(x,t\right)}  &=&
-{\si\left(x,t\right)\over g^2} + iN ~{\rm tr} \left[\langle x,t | 
{1\over i\notpa
-\si} | x,t \rangle \right]= 0\,.
\label{saddle}
\eeqra
In particular, the non-perturbative vacuum of the GN model is 
governed by the simplest large $N$ saddle point of the path integral 
associated with it, where the composite scalar operator 
$\bar\psi\psi$ develops a space-time independent expectation value.  
Thus, the fermions acquire mass 
$m = \Lambda\,\exp\,\left[-{\pi\over Ng^2\left(\Lambda\right)}\right]$
($ \Lambda$ is an ultraviolet cutoff). This mass is associated with the 
dynamical breakdown of the discrete chiral symmetry by the non-perturbative 
vacuum. Associated with this breakdown of the discrete symmetry is 
a kink solution \cite{ccgz,dhn,josh1}, the CCGZ kink mentioned above,
$\sigx = m\,{\rm tanh}(mx)$. It is topology which insures the stability of 
these kinks. The GN model bears also non-topological solitons. These 
non-topological solitons are stabilized dynamically, by trapping fermions 
and releasing binding energy. Since the works of DHN and of Shei, these 
fermion bags were discussed in the literature several other times, using 
alternative methods \cite{others}. For a recent review on these and related 
matters, see \cite{thies}.

The remarkable discovery DHN made was that all the physically 
admissible static, space-dependent solutions of 
(\ref{saddle}), i.e., the static bag configurations in the GN model (the 
CCGZ kink being a non-trivial example of which), 
were {\em reflectionless}. 
That is, the static $\sigx$'s
that solve the saddle point equations of the GN model are such  
that the reflection coefficient of the Dirac equation 
\beq\label{diraceq}
\left[i\notpa-\si (x)\right]\,\psi = 0\,,
\eeq
associated with the GN action, vanishes identically. 

A reflectionless configuration $\sigx$ which supports $K$ 
bound states of the Dirac equation (\ref{diraceq}) is parametrized by a 
a discrete set of $2K$ real parameters 
$0\leq \om_1 < \om_2 < \cdots < \om_K < m\,\,;\,\,
c_1,c_2,\cdots, c_K$, where the $\pm\om_i$ are the bound state energies,
and the $c_i$ determine the asymptotic behavior of the bound state 
wave functions. (Note that the energy spectrum of our one-dimensional 
static Dirac equation cannot be degenerate, and thus no two $\om_i$'s are 
allowed to be equal.) Then, the effective action 
(\ref{effective}), evaluated on this reflectionless $\sigx$ configuration, 
becomes an ordinary function of these parameters. Minus that function, per 
unit time, is the rest mass ${\cal M}(\om_i, c_i)$ of the static lump. 
Thus, the formidable problem of finding the extremal static $\sigx$ 
configurations of (\ref{effective}), is reduced to the mundane problem of 
extremizing the ordinary function ${\cal M}(\om_i, c_i)$.

In this talk we classify the entire spectrum of stable static fermion bags in 
the GN model, thus extending the ``static part'' of the work\cite{dhn} of 
DHN, who considered fermion bags for which the Dirac equation (\ref{diraceq})
had a single bound state. In the narrow space of these pages we can only 
summarize the results. The technical details of our analysis 
will be elaborated elsewhere \cite{periodic}.

First, let us consider the local extrema of ${\cal M}(\om_i, c_i)$. One 
immediate result is that ${\cal M}$ is independent of the parameters $c_i$. 
The $c_i$'s are flat, collective coordinates of the soliton bag.
Consider next the $K$ ordered bound state frequencies 
$0\leq \om_1 < \om_2 < \cdots < \om_K < m\,.$

Assume first that $\om_1>0$. Due to invariance under 
charge-conjugation, the Dirac equation (\ref{diraceq}) has $2K$ bound states  
at $\pm\om_i$. Imagine populating these bound states with fermions. Due to 
Pauli's principle, each energy level (positive or negative) may be occupied 
by up to $N$ fermions, with different flavor indices. Consider 
a given pair of charge conjugate energy levels $\pm\om_i$, which bind 
$N-h_i$ fermions in $-\om_i$ and $p_i$ fermions in $+\om_i$ (with 
$0\leq h_i,p_i \leq N$). We will refer to such a $\pm\om_i$ pair 
as containing $h_i$ holes (or anti-particles) and $p_i$ particles. 
Thus, it contains $n_i=p_i+h_i$ bounded fermions or antifermions in all, 
where, of course, $0\leq n_i \leq 2N$. The total (valence) fermion 
number of such a configuration is obviously $N_{F,val} = p_i-h_i$. It 
turns out that the contribution of these bound states to the total mass 
of the soliton depends only on $n_i$, i.e., we obtain a multiplet of 
states degenerate in energy. By simple counting, for a fixed $n_i$, we find 
that the multiplet contains $(2N)!/n_i!(2N-n_i)!$ states, which we 
immediately identify as an $O(2N)$ anti-symmetrical tensor multiplet of 
rank $n_i$. This multiplet contains states with different numbers of 
particles and antiparticles, hence with different values of the fermion 
number. In fact, it is a straightforward calculation to show that in this 
$O(2N)$ multiplet $-n_i\leq N_{F,val}\leq n_i$ if $n_i\leq N$, and  
$-(2N-n_i)\leq N_{F,val}\leq 2N-n_i$ if $N\leq n_i\leq 2N$. That states with 
different fermion number are degenerate in energy arises here because the 
fermion number $N_F$ is a generator of the $O(2N)$ algebra which commutes 
with the $U(N)$ flavor subgroup, but has non-trivial commutation relations
with the other generators.

It can further be shown\cite{dhn,josh1} that the extremal 
values of the frequencies are $\om_i = m\,\cos 
\left({\pi n_i\over 2 N}\right)$. Thus, each energy level is 
determined by the number of fermions trapped in it. In fact, such a level 
cannot exist {\em unless} it binds fermions. A notable exception is 
a bound state at $\om=0$, if it exists.

A bound state at $\om=0$ is self-charge-conjugate. It arises if and only 
if the extremal $\sigx$ is topologically non-trivial (i.e., a kink or 
an anti-kink). It arises due to topology, and is thus stable, regardless 
of the number $0\leq n_0 \leq N$ of valence fermions it traps. All 
kink states\cite{witten} fall into the (reducible) $2^N$ dimensional 
spinor representation of $O(2N)$. The fermion number eigenvalues of states of 
this multiplet are $-N/2, -N/2 +1, \ldots N/2$, which demonstrate the 
phenomenon of fermion number fractionalization \cite{jr} by the topological 
background.

Consider an extremal $\sigx$ configuration, with $n_i$ 
trapped fermions occupying the pair $\pm\om_i$ ($i=1,\ldots K$), 
as described above. It corresponds to a reducible $O(2N)$ representation made 
of a tensor product of the corresponding antisymmetric $O(2N)$ tensors.  
If this configuration is a kink or an anti-kink, it will have in addition 
a bound state at $\om_0=0$ (with $n_0$ valence fermions), and will add an 
additional factor of the $2^N$ spinorial representation to the multiplet.
It can be shown\cite{dhn,periodic} that the mass of such a 
configuration is $M_{soliton} = 
\sum_{i=1}^K {2Nm\over\pi}\,\sin \left({\pi n_i\over 2 N}\right) + 
{|q|Nm\over\pi}$. Here $q$ is the topological charge ($q=\pm 1$ for kink and 
anti-kink, and $q=0$ in the topologically trivial sector). The sum runs only 
over the non-zero energy levels. Note (for $q=\pm 1$) that the kink mass 
is independent of $n_0$. This extremal configuration corresponds to a 
tensor product of the appropriate antisymmetrical tensors (and also 
the spinorial representation, if $|q|=1$). We verified in \cite{periodic} 
(using resolvent methods) that the expectation value $N_F$ of the fermion 
number operator, in the background of this bag, is simpy 
$N_F = n_0 + \sum_{i=1}^K (p_i - h_i) -{|q|N\over 2}$. The sum part of 
this expression for $N_F$ is the contribution of the valence fermions trapped
in the soliton, and the second part, which arises only in the 
topologically non-trivial sector, is the fractional part. Note that the 
binding energy $B = m\sum_i n_i - M_{soliton}$ increases as each of the 
occupation numbers $n_i$ tends to its maximal possible value $2N$. The more 
fermions the bag traps, the more stable it becomes. This is known in the 
physics of fermion bags as the ``mattress effect''.

The most important question is which of the extremal fermion bags we have 
just described are stable, i.e., true local minima of (\ref{effective}), 
which correspond to eigenstates of the hamiltonian of the GN model (at 
least in the large $N$ limit)? 

Each decay process must conserve, of course, energy and momentum, the 
$O(2N)$ charges (and in particular, the fermion number $N_F$), and the 
topological charge. The list of stable bags, arranged according to their 
$O(2N)$ quantum numbers (the ``periodic table'') is as 
follows\cite{periodic}:\\
\noindent {\it The topologically trivial sector, $q=0$:}\\
{\bf 1)}~Bags with a single bound state, $K=1$, with $\om_1=m\,\cos 
\left({\pi n\over 2 N}\right)\,,~0\leq n_1 \leq 2N$. These are the bags 
discussed in \cite{dhn} (and we shall refer to them as ``DHN bags'').
The mass of such a soliton is $M_n = {2Nm\over\pi}\,\sin \left({\pi n\over 2 N}
\right)$ and its profile is $\sigx = \si(\infty) + 
\kappa\, {\rm tanh}\left[\kappa (x-x_0)\right] - \kappa\, {\rm tanh}\left[
\kappa \left(x-x_0\right) + {1\over 2} \log \left({m+\kappa\over m-\kappa}
\right)\right]\,,$ where $\si(\infty) = \si (-\infty) = \pm m$, 
$\kappa = m\,\sin \left({\pi n\over 2 N}\right) =\sqrt{m^2-\om_1^2}$ and 
$x_0$ is a translational collective mode. It corresponds to an 
antisymmetric tensor representation of $O(2N)$ of rank $n$. Such a bag 
cannot decay into lighter DHN bags, because any such presumed process can be 
shown to violate $N_F$ conservation. Thus, they are stable. 
Note that $M_{n=N} = {2Nm\over\pi} = 2M_{CCGZ}$, and more over, 
that as $n\rightarrow N$ (i.e., $\kappa\rightarrow m$), the profile 
$\sigx$ tends to $\si(\infty) + m\, {\rm tanh}\,(mx) - m\, {\rm tanh}\,
[m(x-R)]\,,~R\rightarrow\infty\,.$ Thus, the configuration 
at $n=N$ is that of infinitely separated CCGZ kink and anti-kink bound at 
threshold. 
\newline
{\bf 2)}~There are no stable, static, topologically trivial bags with 
$K\geq 2$ bound states. (In this case there are always decay channels which 
respect energy and fermion number conservation.)\\
\noindent {\it The topologically non-trivial sector, $q=\pm 1$:}\\
{\bf 3)}~Topological bags with a single bound state, $K=1$. 
These are the CCGZ kinks (or anti-kinks), $\sigx = 
\pm m {\rm tanh}\,\left(m (x-x_0)\right)$. They are the lightest 
topologically non-trivial states. Thus, they are stable. The single bound 
state is at $\om_0=0$, by topology. These kinks form a $2^N$ dimensional 
(reducible) spinorial representation\cite{witten} of $O(2N)$ as was described
above. They trap any number $0\leq n_0\leq N$ of valence fermions. All kink 
states are degenerate in mass at \cite{ccgz} $M_{CCGZ} = {Nm\over\pi}$. 
\newline
{\bf 4)}~Topological bags with $K=2$ bound states. 
There is such a heavier, {\it marginally} stable multiplet of kink 
(or anti-kink) states. By topology, it has a zero mode $\om_0=0$, and in 
addition, a pair of bound states at $\pm \om_1$, with 
$\om_1 = m\,\cos \left({\pi n_1\over 2 N}\right)$. This multiplet is 
reducible: It is the tensor product of the $2^N$ spinorial representation 
and an antisymmetric tensor of rank $n_1$. The mass of this kink is 
$M = {2Nm\over\pi}\,\sin \left({\pi n_1\over 2 N}\right) + M_{CCGZ}$,
and is thus degenerate in mass (in the large $N$ limit) with the sum of 
masses of a CCGZ kink and one of the DHN bags mentioned above. It can be 
shown\cite{periodic} that all presumed decay channels of this heavier kink, 
but two channels, which are allowed by energy conservation and topology, 
violate $N_F$ conservation. The two decay channels which do not 
violate energy and $N_F$ conservation are decays into a CCGZ 
kink plus a DHN bag with either $n=n_1$ or $n=2N-n_1$, which are 
{\it degenerate} in mass with the decaying heavier kink. Thus, these decay 
processes, if not excluded by
other reasons, are at threshold. Thus, the heavier kink is marginally 
stable. However, the profile of this bag is tightly packed:
This kink has two collective coordinates, and in general, its profile
is a rational function of hyperbolic functions. However, for a specific 
choice of the collective coordinates, and for the particular filling 
$n_1=N/3$, the expression for the profile simplifies into $\sigx = \pm m\, 
{\rm tanh}\, (mx/2)\,,$ which is a kink as twice as extended in space as 
the CCGZ kink. Thus, unlike the DHN bag at $n=N$ which we mentioned above, it 
does not have the shape of a configuration of infinitely separated CCGZ kink 
and a DHN bag (whose mass is just the sums of masses of the individual lumps). 
Thus, we conjecture that this heavier kink is a genuine marginally stable 
state in the spectrum. Its connection to the conjectured $S$ matrix of the 
Gross-Neveu model is obscured at this point.
\newline
{\bf 5)}~There are no stable, static, topologically non-trivial bags with 
$K\geq 3$ bound states.

\setcounter{footnote}{0}
\renewcommand{\thefootnote}{\alph{footnote}}

\eject

\nonumsection{Acknowledgements}
\noindent
The organizer of the workshop, Dr. Bordag, should be lauded for a 
job well-done. In addition, I would like to thank him for the kind 
hospitality in Leipzig, during the workshop. Discussions and correspondence 
with M. Moshe and A. Zee are kindly acknowledged. This work was supported in 
part by the Israeli Science Foundation.

\nonumsection{References}

\eject

\end{document}